RESEARCH ARTICLE

# Adaptive optics design for high-energy kW-class multi-slab laser amplifiers

Tomáš Paliesek[1,2], Martin Divoký[1], Jan Pilař[1], Martin Smrž[1], and Tomáš Mocek[1]

[1] *HiLASE Centre, Institute of Physics of the Czech Academy of Sciences, Dolní Břežany, Czech Republic*

[2] *Faculty of Nuclear Sciences and Physical Engineering, Czech Technical University in Prague, Prague, Czech Republic*

**Abstract**

We demonstrate real-time wavefront correction in a high-energy high-average-power DiPOLE100/Bivoj laser using adaptive optics. A bimorph deformable mirror and a Shack–Hartmann wavefront sensor reduced wavefront error 10-fold and improved the Strehl ratio 11-fold. Design aspects such as the deformable mirror actuator geometry, optimal placement and loop frequency are discussed for integration into next-generation high-energy high-average-power lasers.

## 1. Introduction

The DiPOLE laser architecture[1] has become an established laser source in the field of high-energy high-average-power (HEHAP) diode-pumped solid-state lasers (DPSSLs), with three installations in the world[2–4]. The broad adjustability of its output parameters (1–150 J@1030 nm, 2–10 ns, 1–10 Hz) makes it suitable for diverse applications.

The very first system (known as the Bivoj laser system) was delivered in 2015 to the HiLASE Centre (Dolní Břežany, Czech Republic), commissioned in 2016 and in recent years has achieved multiple world records[2,5,6]. It is used mainly for industrial applications (laser shock peening[7] and laser-induced damage threshold testing[8]) and for the research of space-oriented applications (space debris removal[9,10] and spacecraft propulsion).

The second system was supplied to EU-XFEL (Hamburg, Germany) as a laser source for dynamic laser compression experiments at the high-energy-density instrument[11]. The third system is being installed at the Extreme Photonics Applications Center (Oxfordshire, UK) as a pump source for the Ti:sapphire amplifier. The second harmonic frequency (515 nm) will be used to produce 1 PW pulses at a frequency of 10 Hz. In the future, two 100 J amplifiers will be installed here[3]. With recent progress in laser-driven inertial confinement fusion[12,13], the DiPOLE architecture might be considered a potential fusion driver[14].

All of these applications require a high-quality beam in both the near- and far-fields, which is often difficult to maintain because of various degrading effects that decrease the beam quality along the amplification path. Correcting for these effects and achieving the best beam parameters have become an indistinguishable aspect of laser development[15–17].

This paper investigates wavefront aberration correction in the second main amplifier of the Bivoj laser system through three key aspects: aberration characterization, simulation-based adaptive optics (AO) design and experimental testing of the proposed solution. The turbulent cooling introduces random wavefront distortions, while the high average power of the amplifier causes strong static aberration with steep gradients around beam edges that complicate the wavefront sensor (WFS) and wavefront corrector (WFC) placement. Although an AO system was part of the original design, it failed to deliver a flat wavefront at the output of the laser.

Many AO systems have been integrated into low-average-power lasers around the world[18–26]. The low average power (of the order of a few tens of watts maximum) is a consequence of either a very low repetition rate or low energy per pulse. Such output parameters imply rather low thermal loading of gain media, which results in aberration sources coming mainly from optics imperfections or external sources (for example, room air turbulences). Recently, real-time aberration correction in a high-power laser was achieved[27]

Correspondence to: T. Paliesek, HiLASE Centre, Institute of Physics of the Czech Academy of Sciences, Za Radnicí 828, 252 41 Dolní Břežany, Czech Republic. Email: tomas.paliesek@hilase.cz

**Table 1.** High-energy high-average-power laser comparison.

| Laser system | Pulse energy | Rep. rate | Pulse length | Avg. power |
|---|---|---|---|---|
| Bivoj[1,2] (HiLASE) | 146 J | 10 Hz | 2–10 ns | 1460 W |
| HELIA[30,31] (Hamamatsu Photonics) | 253 J | 0.2 Hz | 26 ns | 51 W |
| | 106 J | 10 Hz | | 1060 W |
| DiPOLE–100 Hz[32] (STFC) | 10 J | 100 Hz | 2–10 ns | 1000 W |
| Mercury[33] (LLNL) | 55 J | 10 Hz | 14 ns | 550 W |
| HAPLS[34,35] (ELI Beamlines) | 16 J | 3.3 Hz | 28.6 fs | 53 W |
| HAPLS pump | 97 J | 3.3 Hz | 20 ns | 320 W |

where an AO setup similar to that used in astronomy was employed.

Regarding the DiPOLE laser architecture, four similar laser systems have been developed (see the comparison in Table 1). Higher average power usually means a higher thermal loading of the gain media, higher temperature gradients and therefore higher thermally induced aberrations. Moreover, a higher thermal load requires advanced cooling techniques, which are usually the source of other aberrations that are more random in character. To the best of our knowledge, the only work to date that addresses aberration correction in this type of laser system is Ref. [28]; however, it does not provide information on the temporal dynamics of the aberrations or the performance of AO.

## 2. Bivoj laser system

The Bivoj laser system consists of the front-end followed by two main cryogenically cooled multi-pass amplifiers (Figure 1). The front-end consists of two preamplifiers and generates pulses with energy up to 50 mJ, frequency of 10 Hz, variable pulse shape, variable pulse length (2–10 ns) and adjustable beam profile.

The first main amplifier head (MA1) includes a set of four ytterbium-doped yttrium aluminum garnet (Yb:YAG) gain disks. The pulse passes the gain media seven times using relay-imaging angular multiplexing geometry with a vacuum spatial filter after each pass, as shown in Figure 2. After the third pass, the deformable mirror (DM) compensates aberrations in the feedback loop with the WFS placed at the end of the amplifier. Then the beam transport section expands the beam from 22 mm × 22 mm to 75 mm × 75 mm

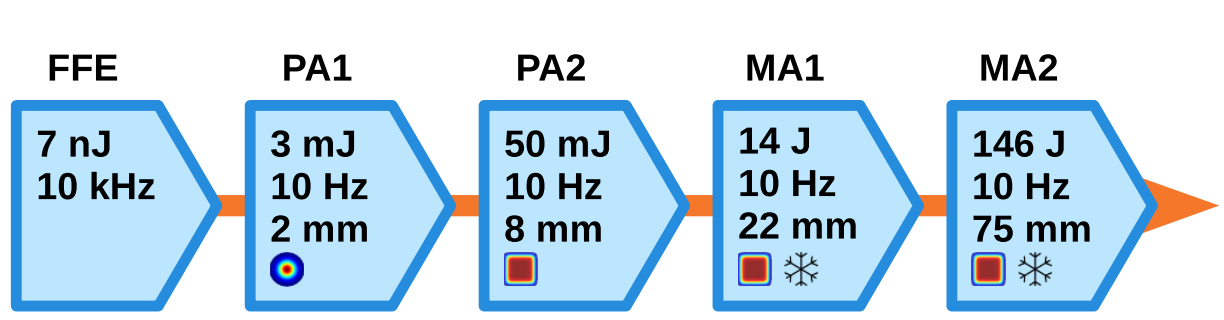

**Figure 1.** Schematic of the individual stages of the Bivoj laser system. FFE, fiber front-end; MA, main amplifier; PA, preamplifier.

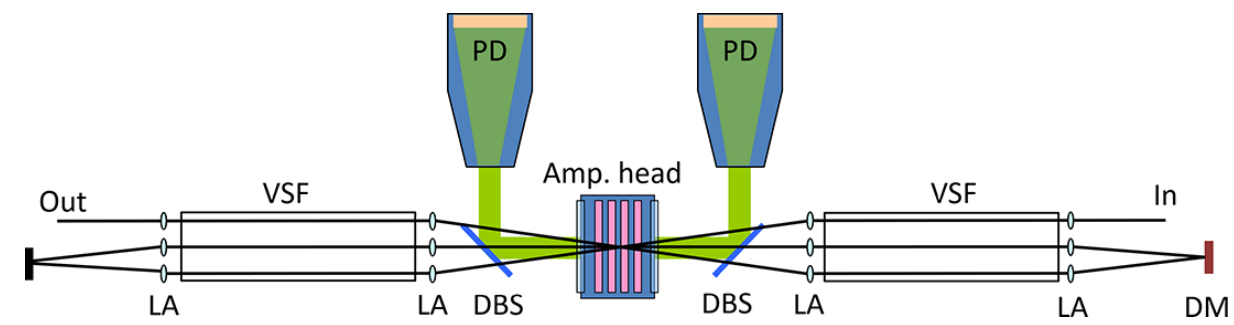

**Figure 2.** Schematic of the multi-pass architecture of the MA1 amplifier[29]. PD, pump diode; DBS, dichroic beam splitter; LA, lens array; VSF, vacuum spatial filter; DM, deformable mirror.

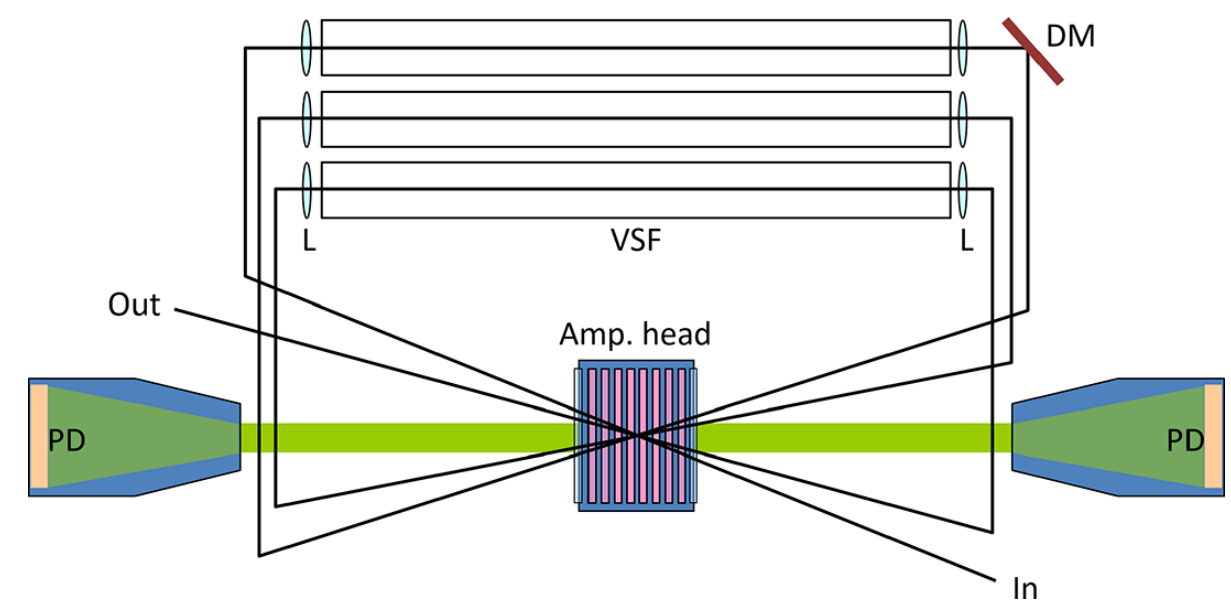

**Figure 3.** Schematic of the MA2 amplifier multi-pass architecture[29]. PD, pump diode; VSF, vacuum spatial filter; L, lens; DM, deformable mirror.

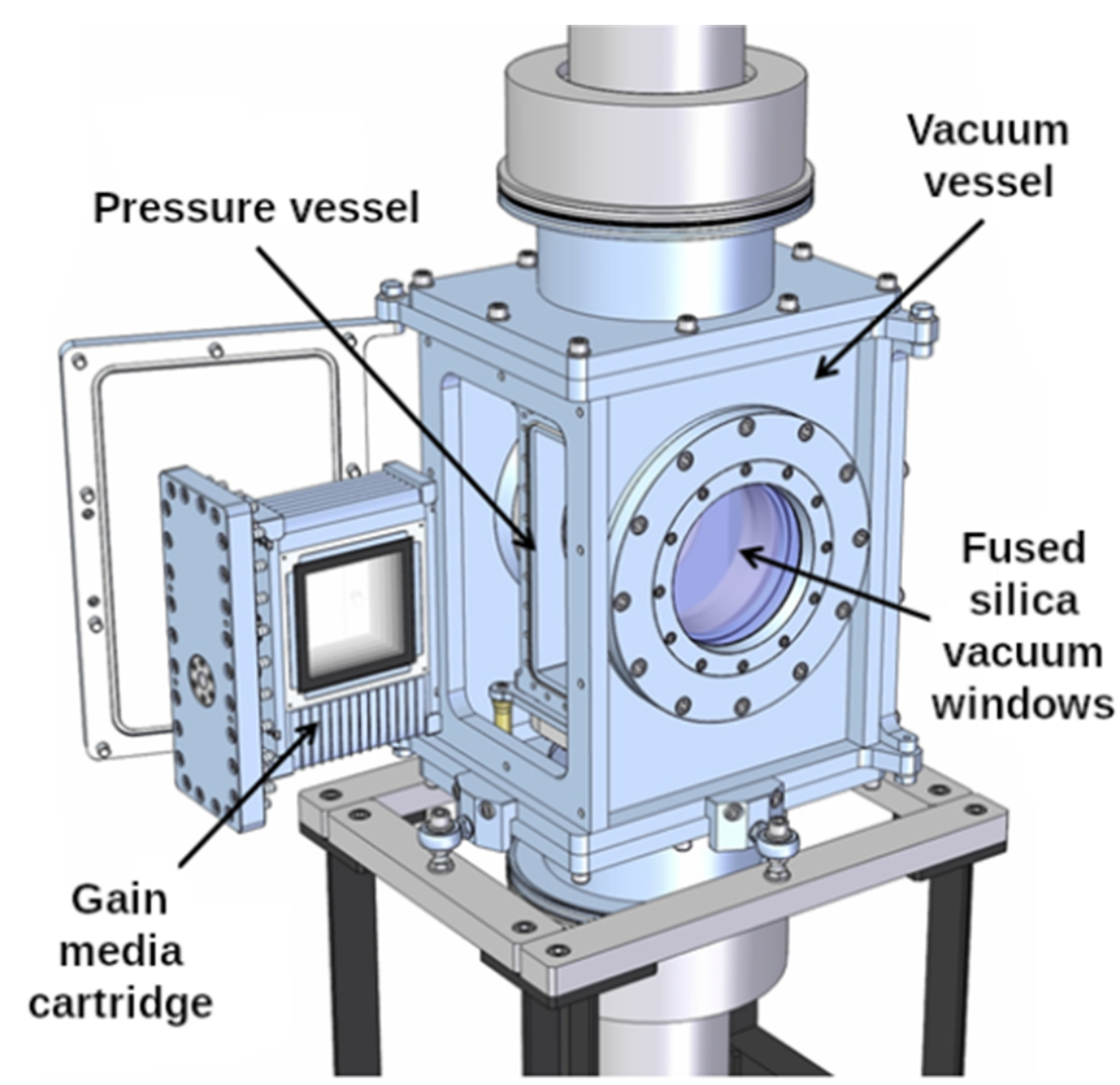

**Figure 4.** MA2 amplifier head model. Reprinted with permission from Ref. [1].

and directs it to the second main amplifier, including optical isolation.

The second main amplifier (MA2) subsequently boosts the pulse energy to 146 J. The circular design shown in Figure 3 features an amplifier head with six square-shaped variably doped Yb:YAG slabs. The cooling helium gas flows through 2 mm wide gaps between slabs (see Figure 4) and the operating temperature is 120–150 K. Two pump units deliver energy from both sides of the head. The beam passes the head four times. Relay imaging on each pass is provided by a Keplerian telescope with 3 m long lenses and a vacuum spatial filter (VSF) with a 3 mm pinhole in the common focal plane to suppress stray light. It is necessary to mention that the relay-imaging distances are not precisely followed and each subsequent head plane drifts slightly from the respective image plane (explained in Section 2.1).

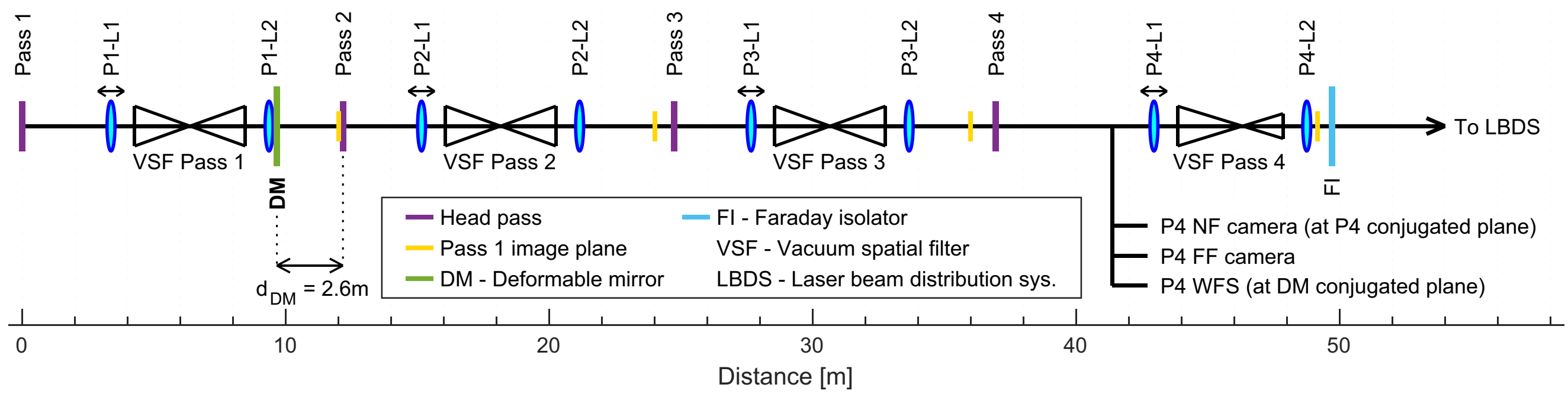

**Figure 5.** Original line scheme of the MA2 amplifier.

After the first VSF, a DM is used to compensate for thermal aberrations within the amplifier in the original design. The DM is controlled by a feedback loop with a WFS that samples mirror leak of the output beam. The DM position and performance are thoroughly explained in the following sections. A more detailed description of the DiPOLE laser architecture can be found in Refs. [1, 36].

Relay imaging is utilized in the whole laser system. The object plane is set by the spatial light modulator (SLM) in the front-end of the system[16] and the image plane is subsequently re-formed by relay telescopes at the following locations:

- the second preamplifier head;
- each head pass of the MA1 amplifier (2× at every pass);
- the MA1 Faraday isolator;
- the first head pass of the MA2 amplifier.

However, this is not followed in the MA2 amplifier, as can be seen in Figure 5. The yellow lines represent the single conjugate planes of the first MA2 head pass. The third head pass is located 74 cm away and the fourth head pass is 95 cm away from the first-pass conjugate plane. The DM is located 2.6 m before the second head pass, and the WFS is placed in the diagnostic line before the fourth VSF in the DM conjugate plane. The thermal lens in the amplifier head is compensated for by shifting the first VSF lens at each pass (marked with horizontal arrows above the lenses in Figure 5).

## 3. Aberration characterization

The aberration sources in HEHAP multi-slab multi-pass amplifiers have already been identified[37–39] and can be divided according to their dynamics as follows.

- Static:
  - heating by pumping;
  - optics defects or imperfections.
- Dynamic:
  - cooling;
  - room air conditioning;
  - head window blow.

Heating of the gain media is caused by two processes: firstly, by non-radiative transitions after absorbing the pump radiation in the central part of the Yb:YAG slabs and, secondly, by amplified spontaneous emission absorption by the Cr:YAG cladding.

The subsequent aberration description refers to operation at the 100 J level with a cooling flow rate of 130 g/s and the gain media temperature of 150 K. The thermally induced defocus is compensated by moving the first telescope lens at each VSF and is equivalent to a thermal lens focal length of 234 m. The thermal equilibrium stabilizes around 30 s after the pump diodes are turned on.

The static component of aberration has a square geometry due to the gain media shape, as can be seen in Figure 6(a). The profiles in Figure 6(c) show that the gradients are steepest near the edges of the beam. Usually, when the amplifier is operated at the 100 J level, the output wavefront has a standard deviation (STD) value of 0.4λ–0.6λ and peak-to-valley (PtV) value of 3λ–4λ (after the defocus is compensated). This results in a Strehl ratio of around 0.15 and a very distorted far-field image, as shown in Figure 6(d).

The dynamic sources of aberration are random in character. The rapid turbulent helium gas flowing in between aerodynamically shaped gain media vanes (Reynolds number >19,000) efficiently cools each gain media face. The MA2 beam path is only partially covered and therefore the room air conditioning also contributes to fluctuations. The amplifier head windows must be under a constant flow of nitrogen gas to avoid condensation of air humidity when the amplifier is cooled down.

The evolution of the wavefront parameters is presented in Figure 7. The data show no periodic changes, and the pulse-to-pulse stability of the STD was within 0.04λ. To further quantify the effects that degrade wavefront quality, the continuous wave (CW) alignment beam (1030 nm, single-mode laser diode) was propagated through the amplifier at various settings, and wavefront evolution over a period of 100 s was measured. The measurements involved changing temperature, cooling helium gas flow and pumping power. The original DM was removed from the amplifier and replaced by a plane mirror.

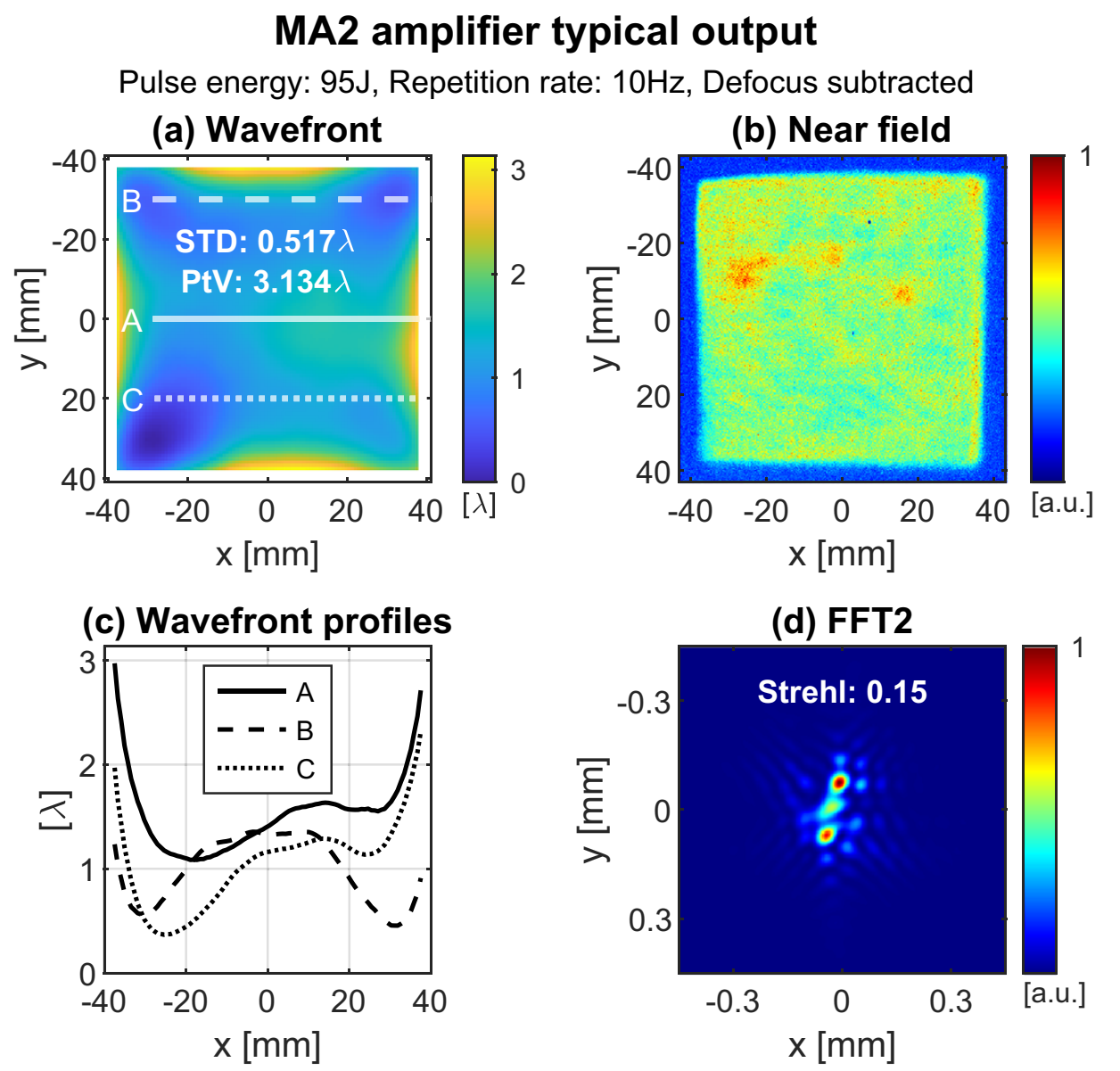

**Figure 6.** MA2 typical output beam and wavefront with subtracted defocus. Line profiles show steep gradients near beam edges.

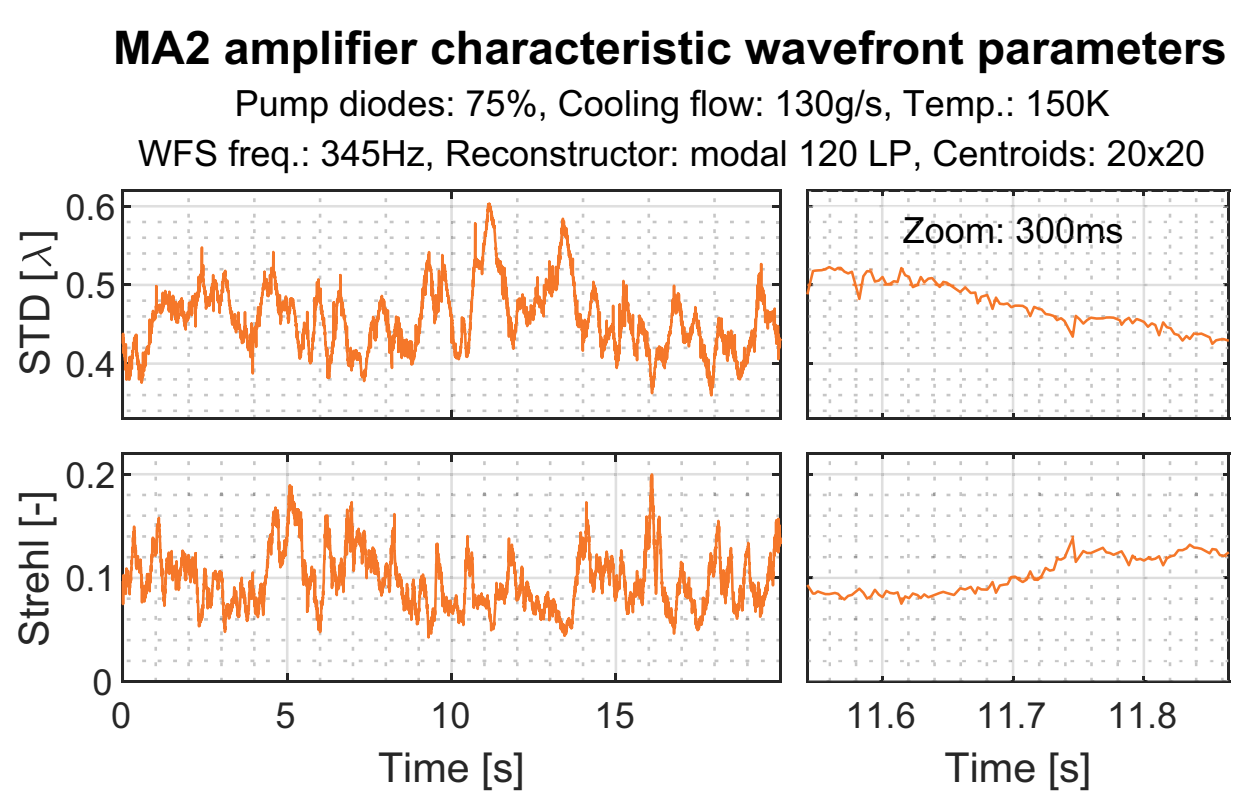

**Figure 7.** Evolution analysis of aberration parameters for the MA2 amplifier.

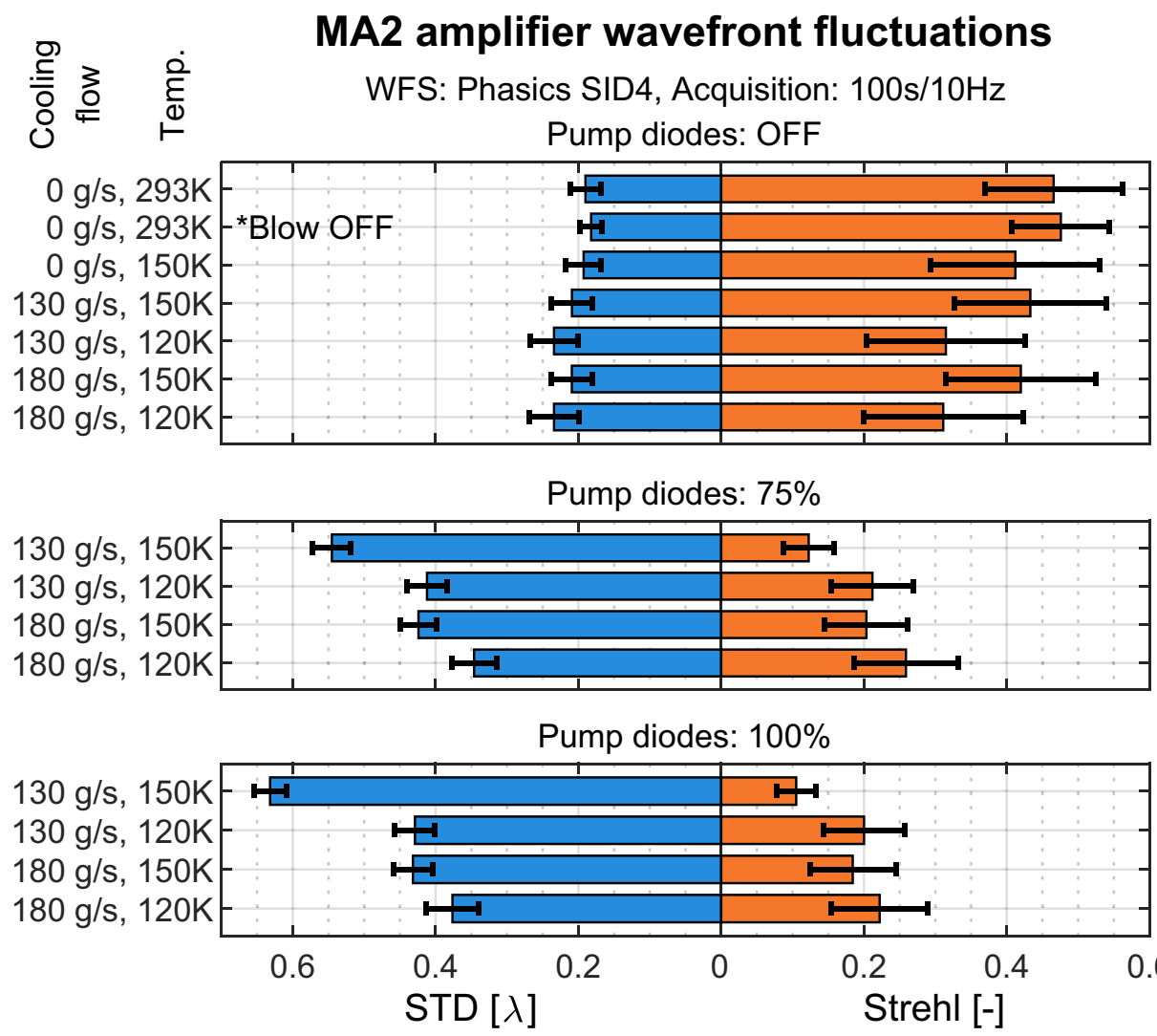

**Figure 8.** MA2 amplifier wavefront STD and Strehl ratio fluctuation analysis for various amplifier settings.

The results can be observed in Figure 8. Simply propagating the beam through the non-cooled amplifier decreases the Strehl ratio to 0.45. The STD of the Strehl ratio is around 50% lower when the head window blow is turned off at room temperature. After the head is cooled down and the cooling flow is turned off, the Strehl ratio decreases even more (~0.4), and the Strehl ratio fluctuations increase by 20%.

A 75% pumping rate simulates operation at an output energy of 100 J, where 25% of the absorbed energy is extracted by the output pulses. The 100% pumping rate corresponds to the operation at the 150 J level, in which the pumping power is increased by approximately 25% above the standard 100 J operation parameters and the head temperature is decreased to 120 K. The pumping power increases the STD values and decreases the Strehl ratio mainly due to thermal loading. The cooling power is increased by increasing the flow rate, which also increases fluctuations. However, a higher helium flow rate reduces the STD and increases the Strehl ratio. Both parameters reach improved values at a temperature of 120 K, which is likely due to the thermo-mechanical properties of YAG – its thermal conductivity increases and its thermo-optic coefficient decreases as the temperature lowers[40].

## 4. Adaptive optics design considerations

The following section analyzes the original AO design and discusses why it was unable to reach a flat wavefront at the output of the amplifier. The subsequent sections propose new requirements for the AO system (DM geometry, AO repetition rate, DM position) using simulations based on the wavefront evolution measurements data.

### 4.1. *Original AO design*

The original amplifier design includes a piston-actuated large-aperture DM, ILAO 135×95 (ISP System, France), equipped with 52 stepper-motor actuators arranged in a hexagonal array with one outer ring, as illustrated in Figure 9. The DM is located 20 cm after the second lens of the first-pass VSF at 45° incidence angle (see Figure 3). The positioning was determined by the size of the mirror, the lack of relay image planes and the low energy fluence after the first head pass. The MA2 output diagnostic line uses the 0° mirror leak before passing the last VSF and reduces the beam to 2 mm × 2 mm. A Phasics SID4 WFS is positioned in the conjugate plane of the DM.

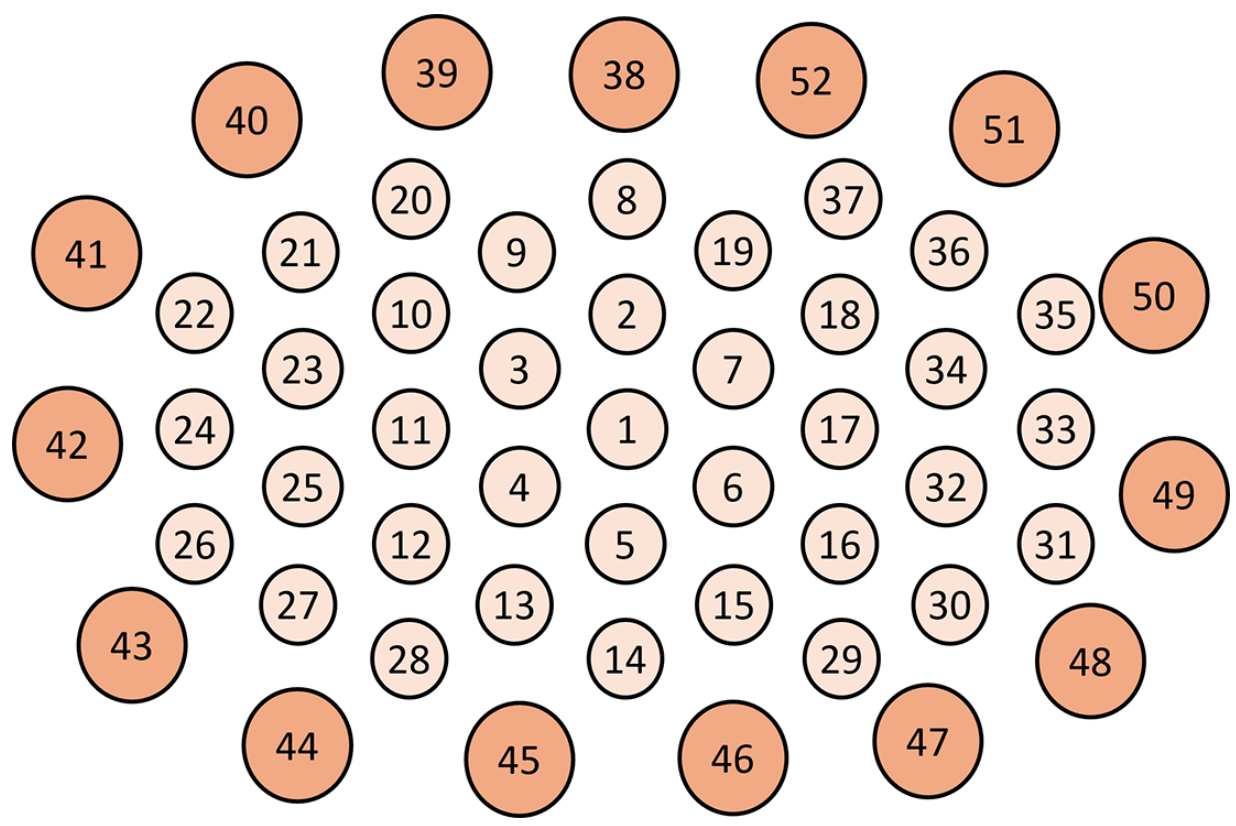

**Figure 9.** Actuator arrangement of the ILAO 135 × 95 DM included in the original AO design.

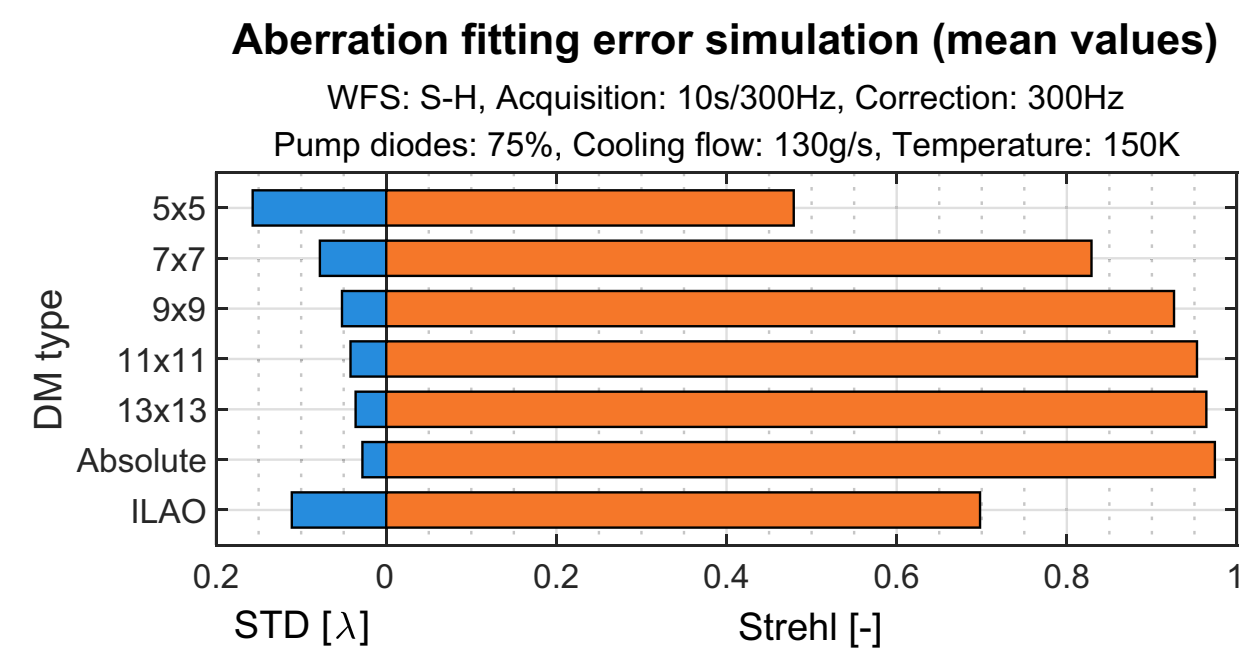

**Figure 10.** Aberration fitting error simulation for various DM geometries. The 300 Hz correction with different DMs was simulated with wavefront evolution data acquired with the Shack–Hartmann WFS.

To evaluate the performance of the DM, a test bench was built. The reflected wavefront of the ILAO DM in its flat state reached an STD of 17 nm over a beam aperture of 75 mm × 75 mm. The DM was also able to replicate the inverse thermal aberration from that in Figure 6(a) up to 58 nm STD error. Despite this performance, the AO loop was unable to converge to a flat wavefront at the amplifier output, as mentioned previously. Eventually, we identified two main reasons for this behavior: the DM location and the long actuator rise time.

After the test bench evaluation, the ILAO DM was returned to the amplifier according to the original design. With cooling turned off and at room temperature (no strong aberrations present in the amplifier head), the ILAO DM was able to achieve a wavefront STD error of 167 nm. Compared to the case without the DM of 190 nm (see Figure 8), this presents a limited improvement. The DM compensates for optics imperfections but is too slow to address the dynamic aberration component. The repetition rate depends on the maximum actuator stroke change and one period typically takes 1–5 s.

When the cooling and pumping are on, the original AO system performance was worse. The best achieved wavefront STD was around 270 nm. In this case, also the DM position in relation to the aberration source plane plays an important role, which is explained in Section 4.4.

### *4.2. Estimation of spatial properties of a DM*

In order to estimate the requirements placed on DM actuators, the performance of several common actuator array distributions was analyzed and compared in a numerical study. The sample DMs corresponding to the selected designs were analyzed in terms of the response functions of all the estimated actuators. The temporal wavefront development subjected to correction was taken from the wavefront evolution measurements within MA2 done with the fast Shack–Hartmann WFS. The spatial distributions of the DM actuator arrays were regular square distributions of 5 × 5, 7 × 7, 9 × 9, 11 × 11, 13 × 13 actuators and the original ILAO DM actuator distribution (52 actuators).

The results for 300 Hz operation are shown in Figure 10 together with the absolute correction result, which represents a result of the direct correction application (correction is applied without being composed of DM eigenmodes) and therefore presents a limit of the correction given only by the limited temporal dynamics and correction gain of the estimated AO system. From the results, it can be seen that in order to address the expected wavefront aberrations, the square arrays perform better than the ILAO DM geometry and that the required number of actuators should be at least 7 × 7.

### *4.3. AO repetition rate*

To determine what AO repetition rate is necessary to achieve a flat wavefront, the wavefront evolution at the MA2 amplifier output was measured using a high-speed Shack–Hartmann WFS, and correction simulation was performed to investigate how varying the loop frequency affects the system's ability to compensate for dynamic wavefront distortions.

The correction was simulated on the wavefront evolution acquired at a frame rate of 300 Hz and with the following amplifier settings: 75% pumping, a cooling flow rate of 130 g/s and a temperature of 150 K. The influence functions (i.e., the responses of individual actuators) of the 8 × 8-actuator square DM were used to apply the correction by subtraction. These influence functions were chosen because this mirror was available for testing at our facility. The correction was computed only at specific frames, depending on the correction frequency, and was applied by subtracting it from the current frame and all subsequent frames. The correction accounted for loop gain but did not consider actuators' limited ranges.

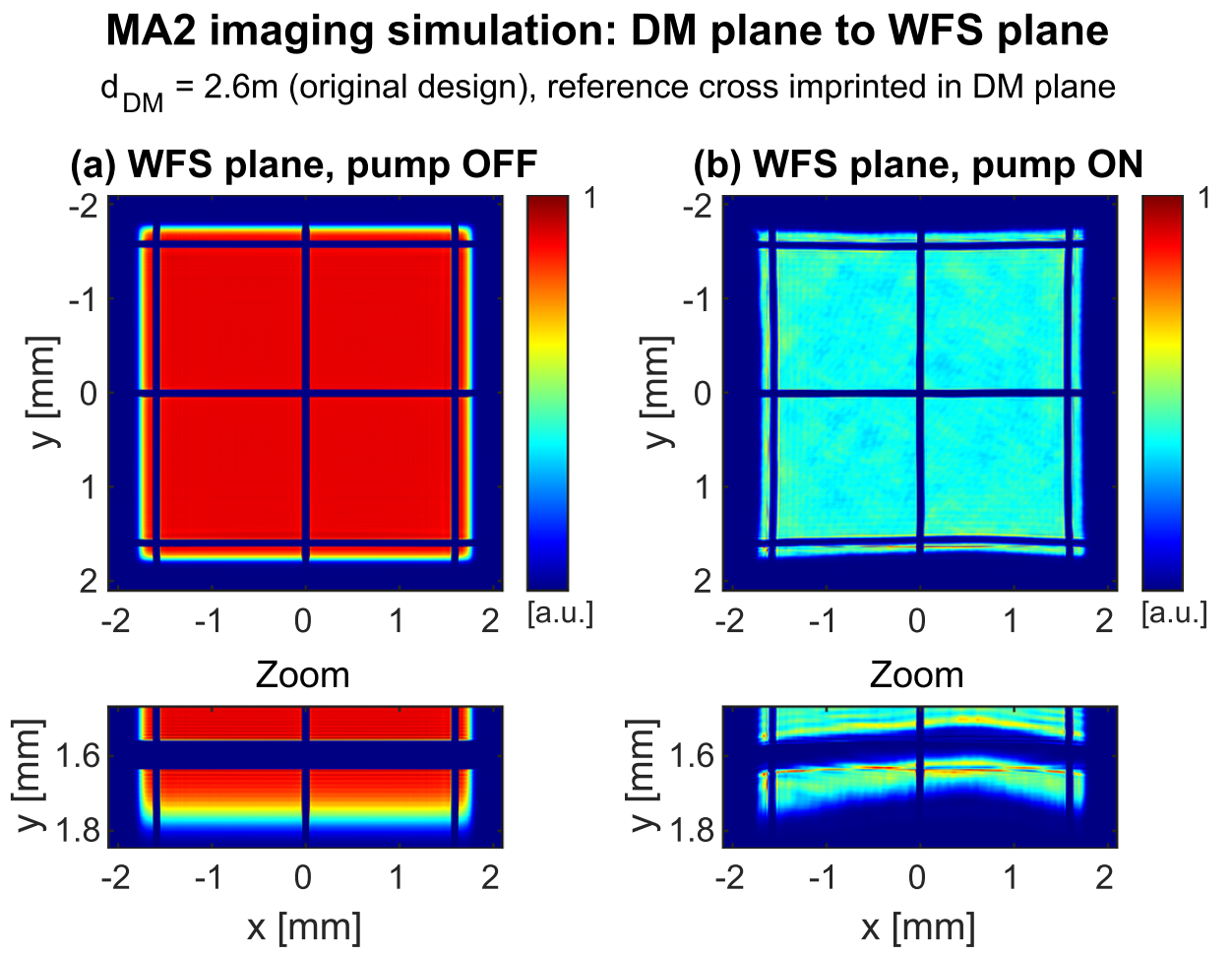

**Figure 11.** Simulation of imaging the DM plane onto the WFS plane in the MA2 amplifier. A beam with a reference cross imprinted in the DM plane is propagated through the MA2 model. The simulation scheme is shown in Figure 5. Wavefront aberrations imposed at each head pass simulate pumping. The imaging is distorted in the presence of strong wavefront gradients that are not conjugate with the DM plane.

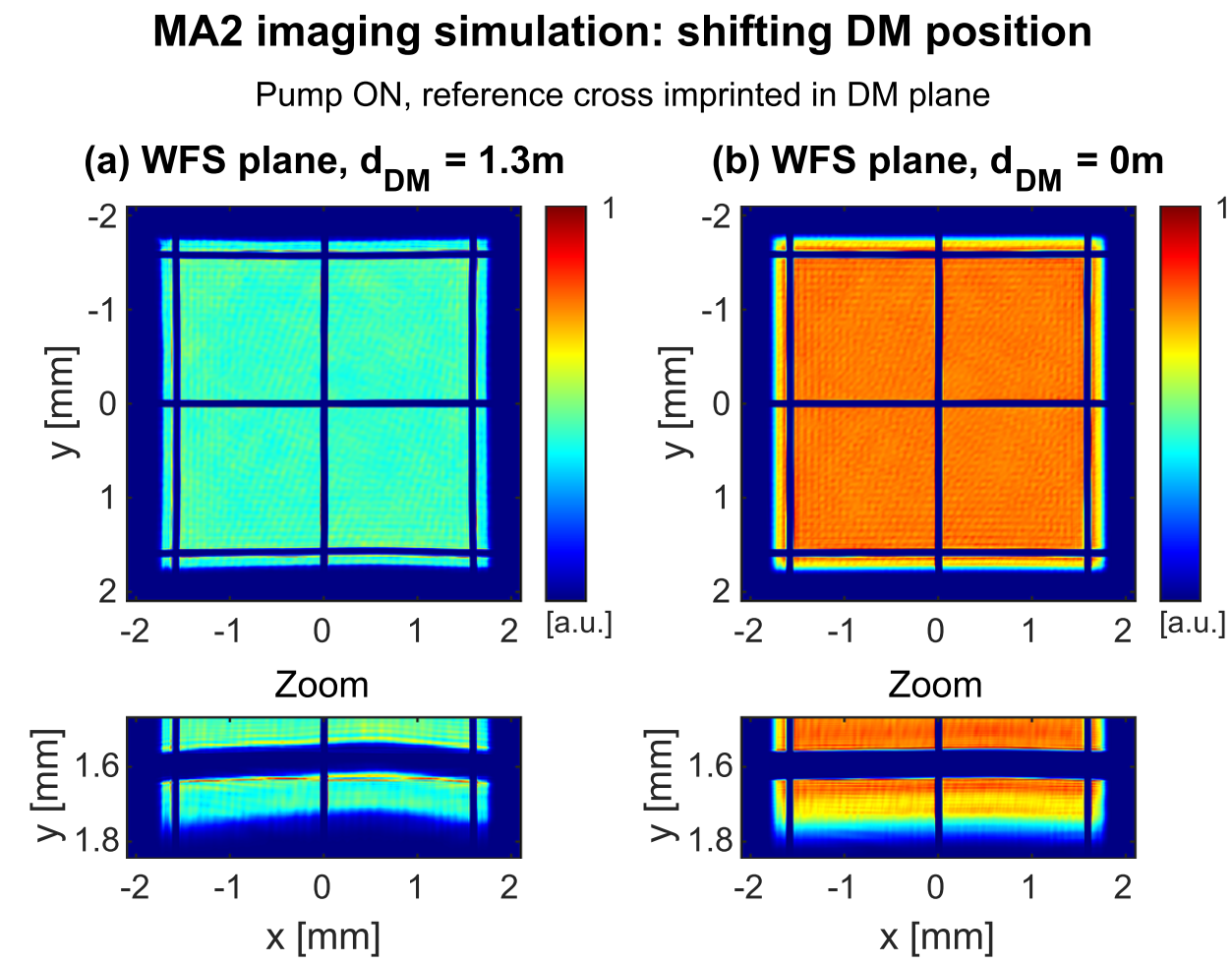

**Figure 12.** Simulation of imaging the DM plane onto the WFS plane in the MA2 amplifier, with the DM moved closer to the head (aberration source plane). The distance $d_{\mathrm{DM}}$ is defined in Figure 5, which also shows the simulation scheme. The imaging is distorted due to strong wavefront gradients that are not conjugate with the DM. The closer the DM is to the head, the less distortion appears in the beam profile.

Depending on the loop gain and the number of eigenmodes used in the simulation, to achieve a mean Strehl value higher than 0.8, the required AO repetition rate of 20–50 Hz was necessary and at least full 300 Hz was required to reach a mean Strehl ratio higher than 0.9. These results were confirmed in the experiment, where the DM at the output of the system was used to compensate for aberrations at various repetition rates (see Section 5 and Figure 17 detailed later).

### *4.4. DM location*

The inability of the original AO design to converge to a flat wavefront led to investigation of the DM position in the MA2 amplifier. The first aspect is the imaging in the strongly perturbed environment of the laser amplifier, discussed in Section 4.4.1. The second aspect is relative positioning of the DM and the main aberration source, explained in Section 4.4.2.

#### *4.4.1. Imaging the DM plane to the WFS plane*

AO systems typically position the WFC and WFS at planes within the optical system that are conjugate to its exit pupil, as this conjugation is crucial for proper AO operation. In the MA2 amplifier case, the DM is conjugate to the WFS according to Figure 5 in the original design. However, they are not conjugate with the aberration source planes (head passes) and these aberrations strongly distort imaging from the DM onto the WFS, as can be seen from the following beam propagation simulation.

Let us now consider imaging the DM plane to the WFS plane in the diagnostic line, according to the original design presented in Figure 5. By simulating the beam propagation in the amplifier, we found that the strong thermally induced aberrations – mainly the strong gradients near the edges of the beam – severely distort the imaging. The reference cross imprinted in the DM plane is intentionally moved near the edges of the beam, where the wavefront gradients are the steepest.

When pumping is on, the straight lines on the DM are no longer straight when imaged onto the WFS plane, as can be seen in Figure 11. Therefore, the actuators' responses seen by the WFS are also deformed. The propagation model based on the angular spectrum method[41,42] included free-space propagation, lenses and aberrations at each head pass.

The simulation allowed one to place an artificial DM plane anywhere in the MA2 amplification path. Figure 12(a) shows less image distortion when the DM plane is placed closer to the second head pass. Nevertheless, even when $d_{\mathrm{DM}} = 0$ m (Figure 12(b)) there are still some scintillation artifacts around the beam edges that are probably caused by the fact that the individual head passes are not perfectly conjugate, as was mentioned before. Following the simulation result, it would be beneficial to position the DM in a head conjugate plane. This is the case of the MA1 amplifier, where before and after each head pass, the image plane is reproduced and the DM is placed in the head conjugate plane after the third head pass (see Figure 2). Unfortunately, this is not possible in the current MA2 setup as there are no image planes available for DM installation. An amplifier redesign would be in order.

#### *4.4.2. Validity of geometric optics*

Another point of view on this issue is whether the geometric optics (GO) validity is followed in the original MA2 amplifier layout. We will use the description given in Ref. [43], which explains the validity of GO for AO for astronomical telescopes. This condition generally holds when diffraction is insufficient to produce significant intensity changes (scintillations) at the telescope pupil, which is often the case for aberration caused by weak turbulence. If this weak turbulence condition is maintained, then light rays emanating from a single-point source can be corrected by placing the WFC at any point along the ray path. If not, interference effects become visible and the resulting phase and intensity variations cannot be completely removed.

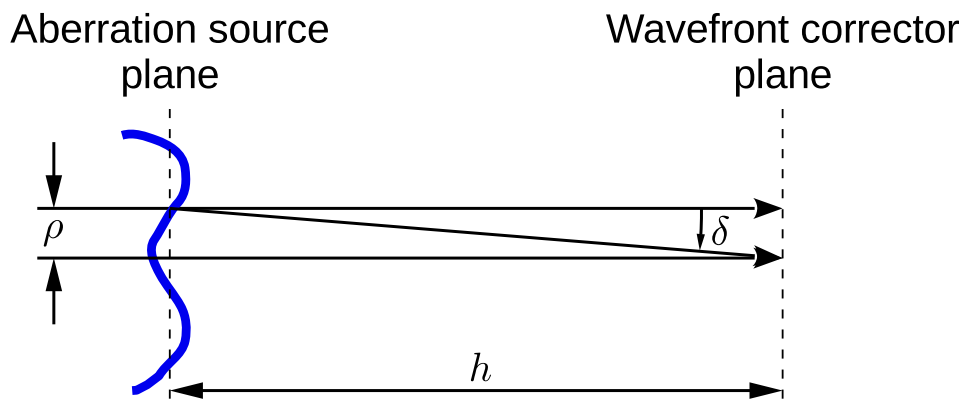

**Figure 13.** Aberrated beams interfere at the wavefront corrector when $h > \rho/\delta = \rho^2/\lambda$. Inspired by Ref. [43].

This poses limits on where the WFC can be positioned, a topic that has been widely studied in the field of multi-conjugate AO for astronomy[44,45] and conjugate AO for microscopy[46,47]. The main idea of this concept is to place the WFCs in planes conjugate to the aberration source layers and compensate for each aberration layer independently. This arrangement preserves the validity of the GO for wavefront correction and restores the spatial relation between the aberration layer and the WFC[43].

Light incident on the aberration layer (see Figure 13) is diffracted at an angle $\delta$ equal to $\lambda/\rho$, where $\lambda$ is the mean wavelength and $\rho$ is the aberration layer scale size. Interference will occur when diffracted rays travel a distance comparable to the following:

$$h = \rho/\delta = \rho^2/\lambda. \tag{1}$$

If the propagation distance between the aberration layer and the WFC is less than $h$, the diffraction effects may be neglected and the wavefront aberrations can be compensated by placing the WFC at any location in the ray path. However, if the aberration is stronger, then the diffracted rays are displaced and produce interference effects. In astronomy, the characteristic size is given by Fried's parameter $r_0$, which is the diameter over which the wavefront aberration has a mean square value of 1 rad$^2$ at a wavelength of 0.5 μm.

Applying this description to the beam in the original DM plane, it is found that the area where the mean square wavefront error (MSWE) reaches 1 rad$^2$ is as small as 7 mm, which gives $h < 47.6$ m according to Equation (1). This condition is satisfied, since the original DM plane is approximately 2.6 m away from the first head pass conjugate plane, indicating that no scintillation occurs in the original DM plane. In contrast, when the same description is applied to the typical output wavefront of the MA2 amplifier, the diameter of the area where the MSWE reaches 1 rad$^2$ differs from only 1.2 mm near the edge to 20 mm in the middle of the beam. Substituting the minimum value into Equation (1) gives $h < 1.4$ m, which represents the distance below which the validity of the GO is maintained if the WFC is placed at the MA2 output.

**Table 2.** MA2 amplifier AO design summary.

| | |
|---|---|
| AO repetition rate | |
| Mean Strehl > 0.8 | 20–40 Hz |
| Mean Strehl > 0.9 | 300 Hz |
| DM distance from aberration source (conjugate) plane | |
| DM at the output | <1.4 m |
| DM inside the amp | 0 m (DM in conjugate plane) |
| DM geometry | |
| Geometry | Square |
| Number of actuators | min. 7 × 7 |

### *4.5. AO design conclusions*

Table 2 summarizes the findings of the previous paragraphs. The required repetition rate implies using a piezoelectric DM with kHz bandwidth featuring square geometry and a high frame rate WFS. The DM in the amplifier should be positioned as close as possible to the aberration source (conjugate) plane to mitigate the DM to WFS imaging distortion. If placed at the MA2 output, the distance from the aberration source conjugate plane should be less than 1.4 m.

## 5. Experimental adaptive optics performance

Based on all considerations, simulations and space limitations, it was decided to place the DM at the second head pass conjugate plane at the amplifier output and test the aberration correction using the CW alignment beam that allows one to achieve a high WFS frame rate. The 8 × 8 bimorph DM with clear aperture of 22 mm × 22 mm was used. This DM was originally part of a different laser system and, at the time of testing, it was available to use at our facility. The WFS was a Shack–Hartmann, capable of a 500 Hz frame rate when the aperture is cropped to 100 centroids. The control software as well as the DM and WFC was supplied by Dynamic Optics. The amplifier parameters were set to reproduce the thermal loading during 100 J operation (pumping power: 75%, cooling flow rate: 130 g/s, temperature: 150 K).

The MA2 output laser beam was delivered by the laser beam distribution system (LBDS) to the testing setup. The 75 mm × 75 mm beam was demagnified by a telescope and imaged onto the DM, which was placed in the second head pass conjugate plane. The following telescope re-imaged the DM onto the WFS. Figure 14 illustrates the experimental arrangement. The beam covered 20 × 20 microlenses on the WFS, which allowed us to operate the WFS at a frame rate of 345 Hz.

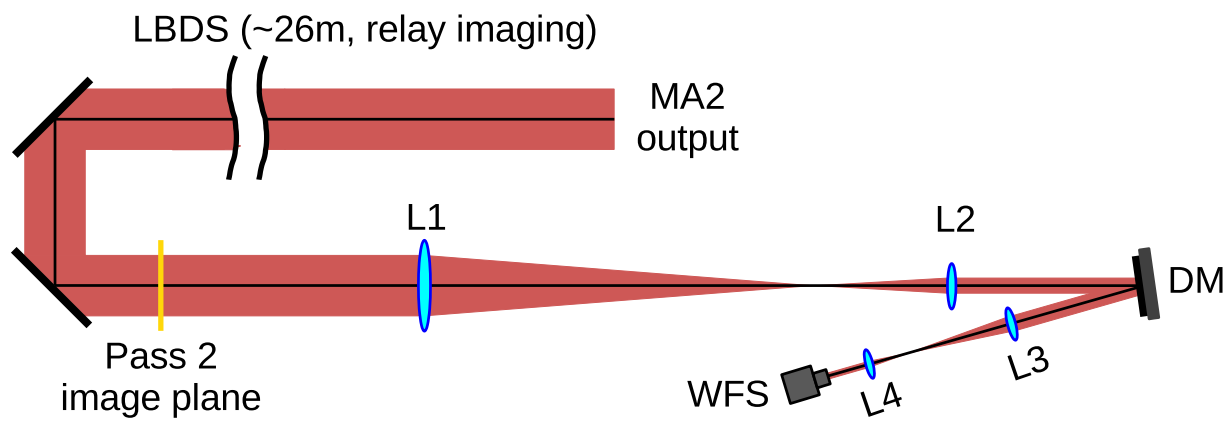

**Figure 14.** Test experiment scheme. The DM, WFS and second head pass plane are all conjugate planes. LBDS, laser beam distribution system; L1–L4, lenses.

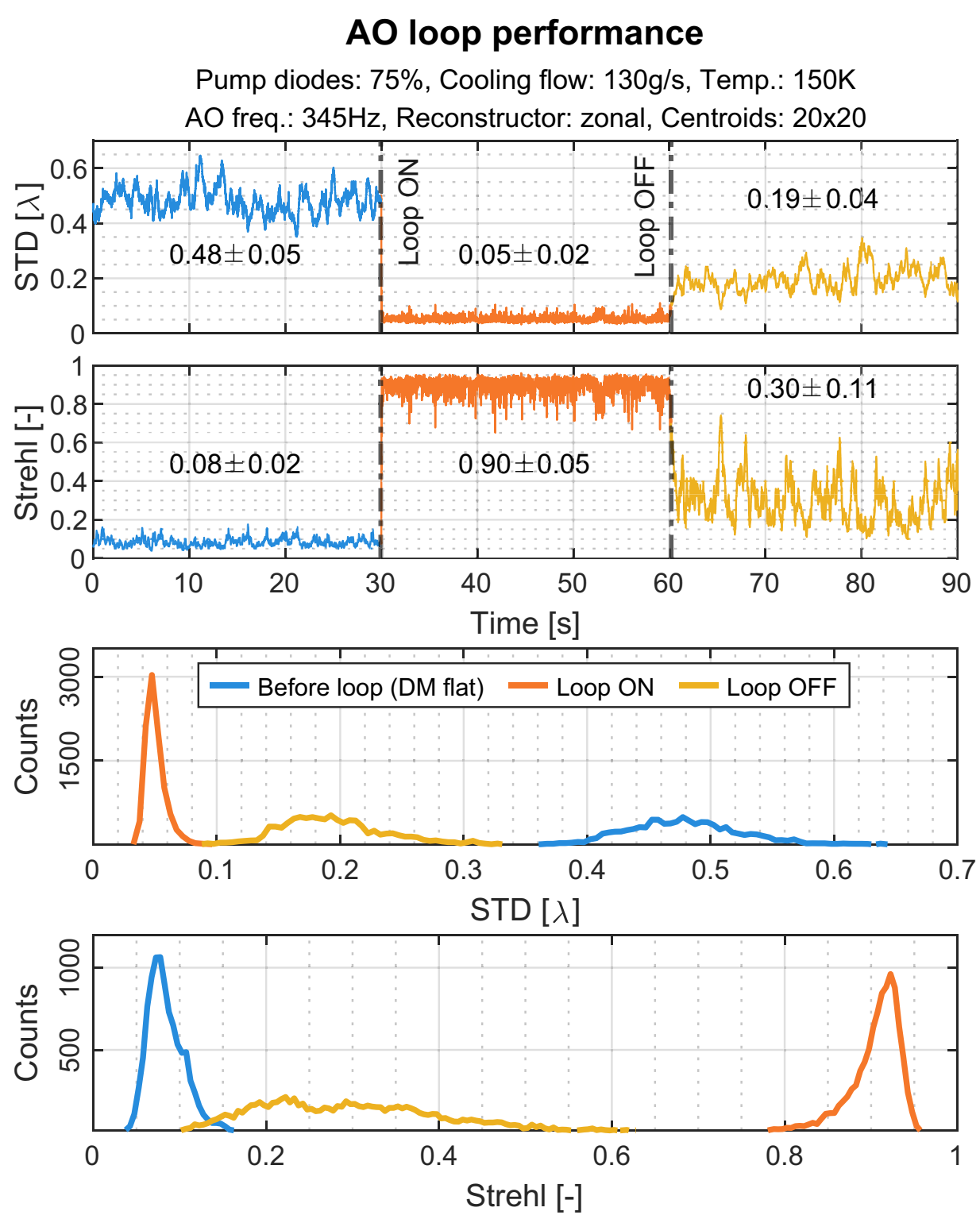

**Figure 15.** Adaptive optics setup performance at 345 Hz. The first two line plots show evolution of the STD and Strehl ratio, while the last two plots present the STD and Strehl ratio histograms.

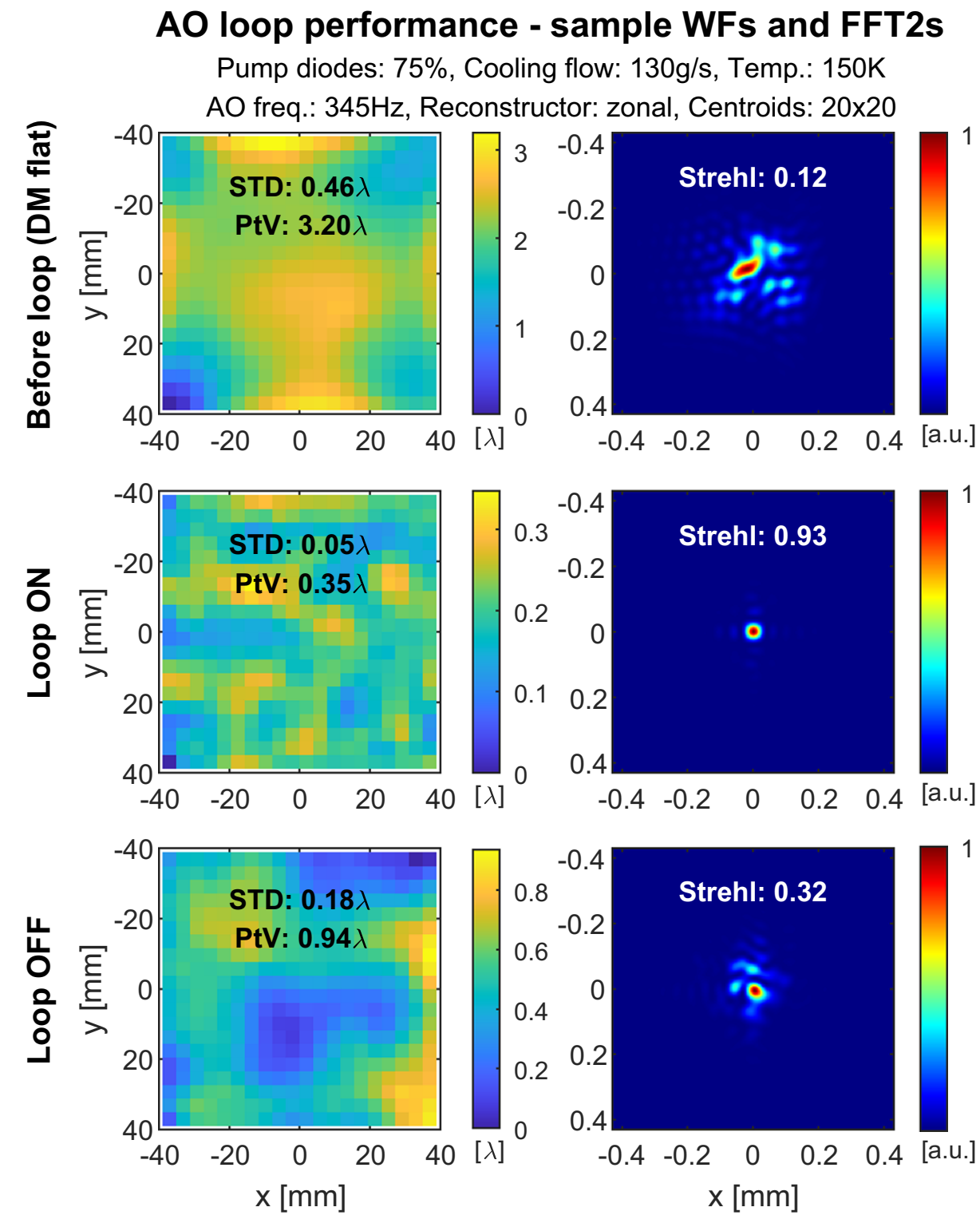

**Figure 16.** Sample wavefronts for individual stages of the AO performance shown in Figure 15.

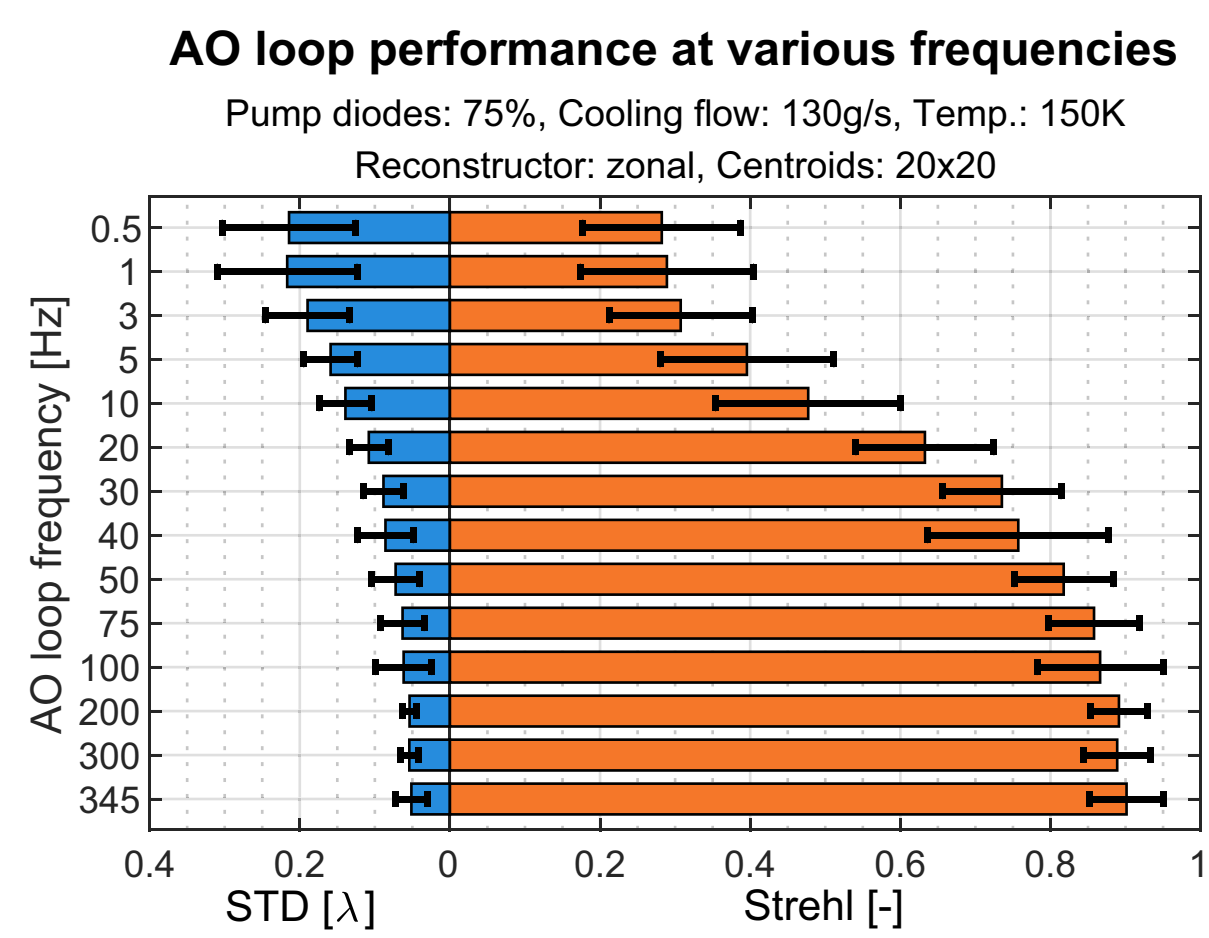

**Figure 17.** AO setup performance at various loop frequencies. Mean values when the AO loop was on are displayed.

Initially, the actuators were set to their zero positions and, after 30 s, the AO loop was started. The loop settling time to reach the Strehl ratio of 0.9 was 62 ms (21 iterations) at a repetition rate of 345 Hz. After another 30 s, the AO loop was stopped to observe aberration evolution with static correction applied at the DM. The performance is presented in plots in Figure 15. The wavefront STD was improved by 10× and the Strehl ratio by 11× when the loop was turned on. After stopping the loop, the static correction improved the Strehl ratio by only 4× and the STD by 2.5×, mainly because of the dynamic nature of the aberrations. The sample wavefronts for each measurement section are shown in Figure 16 together with individual far-field distributions.

Figure 17 shows the AO performance at different repetition rates. The repetition rate necessary to achieve the mean Strehl ratio of 0.8 was 50 Hz. The repetition rate of 345 Hz allowed us to exceed the Strehl ratio of 0.9. The results are in good agreement with the simulations results. It should be noted that the AO loop did not take into account changes of pointing because tilt stability was influenced by the long beam path through the LBDS to the testing setup area and therefore does not correspond to the pointing stability at the output of the amplifier.

## 6. Discussion

The results show that aberration correction in MA2 is possible. However, the DM will have to be placed as close as possible to the head conjugate plane either before the beam enters MA2, inside MA2 or less than 1.4 m away from the head conjugate plane at the output. In addition, a probe beam will have to be added to the amplifier chain, allowing the AO system to work at high frame rates.

The first option would require a DM for a 22 mm × 22 mm beam that withstands a fluence of 3.2 $J/cm^2$ (which might be challenging for large apertures[48]) or redesigning the beam transport section that connects the MA1 and MA2 amplifiers, while maintaining the current MA1 output beam size and image plane, which are often used separately. If the DM was placed inside MA2, MA2 redesign would be required, as well as a large-aperture DM (for a 75 mm × 75 mm beam) that could operate in the kHz bandwidth. The last option would also require the same type of DM as option 2 except with a higher damage threshold specification ($> 2\ J/cm^2$).

## 7. Conclusion

In summary, this paper investigates the design of the AO system for the HEHAP Bivoj laser with possible impact on all DiPOLE100 architecture lasers worldwide. In order to determine the necessary parameters of the AO system, aberration characterization was carried out. The aberration composition is unique because high average power implies a strong static component, and turbulent cooling adds a strong random character. To the best of our knowledge, no other works on this type of aberration correction have been published. What makes it more difficult is that the presence of strong wavefront gradients restricts positioning of the DM.

The test experiment showed a successful aberration correction at the output of the MA2 amplifier. The AO setup used the 8 × 8-actuator bimorph DM, a Shack–Hartmann WFS and PhotoLoop control software (Dynamic Optics, Italy). The wavefront STD was improved 10× and the Strehl ratio 11× when the AO loop was turned on with a repetition rate of 345 Hz. However, due to the high AO repetition rate needed, the experiment was performed at low output energy, only with the CW alignment beam. The position of the DM was discussed throughout the paper from the point of view of conjugation with aberration source plane and also from the point of view of original design limitations.

It was found that the original design with a piston-actuated large-aperture DM would not be able to achieve acceptable wavefront parameters even though it was located in a more suitable position. The repetition rate of the mirror would limit the correction performance because of the dynamic nature of the aberration.

An important contribution of this work is for the new DiPOLE-100Hz[32] laser system that is currently being evaluated in our center. The thermal loading and the cooling-induced aberration will be comparable, but the beam aperture will be 22 mm × 22 mm, making the wavefront gradients even stronger.

## Acknowledgements

This work was co-funded by the European Union and the state budget of the Czech Republic under project LasApp CZ.02.01.01/00/22_008/0004573.

## Data availability

The data that support the findings of this study are openly available in the Zenodo repository at https://doi.org/10.5281/zenodo.16612166.